\documentclass[11pt]{article}
\usepackage{geometry}
\usepackage{graphicx}
\usepackage{amssymb}
\usepackage{amsmath}
\usepackage{epstopdf}
\newcommand{\be}{\begin{equation}}
\newcommand{\ee}{\end{equation}}
\newcommand{\bea}{\begin{eqnarray}}
\newcommand{\eea}{\end{eqnarray}}
\newcommand{\h}{\hspace{0.04 cm}}

\newcommand{\n}{\nonumber}
\title{Analogue Special and General Relativity by Optical Multilayer Thin Films:\\ The Rindler Space Case}
\author{Sh. Dehdashti$^{1}$, R. Roknizadeh$^{1}$, A. Mahdifar$^2$\\
\small{ 1. Department of Physics, Quantum Optics Group, University of Isfahan, Isfahan, Iran}\\
\small{2. Science Department ,  Shahrekord University,Shahrekord}}
\begin{document}
\maketitle
\begin{abstract}
In this paper, to obtain an analogy between the curved spaces and
the linear optics, we expand the idea of Ref.\cite{mon5.5,mon5} to the
multilayer films. We 
investigate effects of thickness and index of refraction of the
films on the Lorentzian transformations. In addition, by using
the multilayer films, we suggest very simple experimental set-up
which can serve as an analogue computer for testing special
relativity. Finally, we draw an analogy between the Rindler
space, as an example of the curved spaces, and a suitable
multilayer film.
\end{abstract}
 {  \bf keywords}: Multi-layers optics, Rindler space, Spinor map, Transfer Matrix.

 \section{Introduction}\label{s1}

In recent years, analogies have taken on a major role in physics
and mathematics \cite{vis,leon1,leo}. They provide the new ways of
looking at problems that permit cross-fertilization of ideas
among different branches of science. Analogy methods have 
practical worth from two point of views:
 First, they provide new light on
perplexing theoretical questions \cite{vis,leon1}. Second,
some of them are interesting for experimental set-up \cite{vis}. \\
\indent In this contribution we use the method developed in
 \cite{mon5.5,mon5}, in a  multilayer stack to show how we can 
 set up an experimental model for special relativity and Rindler 
 space as a special case of general relativity. \\
\indent In the present section we review some of the relevant 
results achieved in \cite{mon5.5,mon5}.
 As it is well known,  $ SL(2, \mathbb{C}) $
group  is locally isomorphic (or a 2 to 1 homomorphic) to Lorentz group, $ SO(1,3)$. 
In fact, the transfer matrix, used in the classical optics for description 
of the  interaction of  light and matter,  
can be considered as an element of $SL(2,\mathbb{C})$
(see \cite{mon1} and references therein).\\
 \indent In ref. \cite{mon5}, authors gave a geometrical interpretation for the transfer
matrix of an absorbing system. In fact,
they considered an isolated absorbing film embedded between two
identical ambient $I$ and substrate $II$ media. A monochromatic,
linear polarized plane wave falls from the ambient, as well as
 a plane wave of the same frequency and polarization falls from the
substrate. If the amplitudes of the field  are characterized
by a column vector,
 \begin{equation}
  \textbf{E}=\left( \begin{array}{c}E_{-}\\ E_{+}\end{array} \right)
 \end{equation}
where $E_{-}$ ($E_{+}$) is denoted backward
(forward)-travelling
 plane wave, and the amplitudes of the ambient and
substrate  are related by
 \begin{equation}\label{0.1}
  \textbf{E}_{I}=\textbf{M}\ \textbf{E}_{II},
 \end{equation}
where $\textbf{M}$ is the transfer matrix and is defined as,
 \begin{equation}
  \textbf{M}=\left( \begin{array}{cc}
  \frac{T^{2}-R^{2}}{T} & \frac{R}{T}\\ -\frac{R}{T} &
  \frac{1}{T}
  \end{array} \right),
 \end{equation}
and $T$ and $R$ are the transmission and reflection coefficients,
respectively.\\
\indent By defining the coherency  matrix \cite{bor} as
 \begin{equation} \label{0.3}
  \mathcal{E}=\left( \begin{array}{cc}
  |E_{-}|^{2} & E_{-}^{*} E_{+}\\ E_{-} E_{+}^{*} & |E_{+}|^{2}
  \end{array} \right),
 \end{equation}
for both $I$ and $II$,  it is easy to show that,
\begin{equation} \label{0.4}
\mathcal{E}_{I}= M\ \mathcal{E}_{II}\ M^{\dagger}.
\end{equation}
It is useful to identify  the variables $x^{\mu}$, 
\begin{equation} \label{0.5}
x^{\mu}=\frac{1}{2} Tr(\mathcal{E} \sigma^{\mu}),
\end{equation}
as the coordinates of a $(1+3)$-dimensional Minkowski space, in which
$\sigma^{0}= \mathbb{I}$ (the identity) and $\sigma^{i}$'s are  the
standard Pauli matrices. The conjugation (\ref{0.4}) induces a transformation on the
variables $ x^{\mu}$ of the form
\begin{equation}\label{2.5}
x_{I}^{\mu}= T_{\nu}^{\mu}(M)\ x_{II}^{\nu},
\end{equation}
where,
\begin{equation}\label{2.6}
T_{\nu}^{\mu}(M)= \frac{1}{2} Tr(\sigma^{\mu} M \sigma_{\nu}
M^{\dagger}).
\end{equation}
It is easy to see that the transformation $T(M)$ is a Lorentz
transformation \cite{neber} which is generated by  $M$ and $-M$. 
Thus it is a $2 \rightarrow 1$ homomorphism map between $SL(2, \mathbb{C}) $ 
and $ SO(1,3) $.\\

\indent We use the model which is developed in \cite{mon5.5,mon5} , 
and consider a monochromatic, linearly
polarized plane electromagnetic wave incident on a 
non-absorption multilayer thin film. In this model   
basis coordinates are
defined by the components of the electromagnetic wave, i.e., $E$ and
$H$. \\

\indent The organization of this paper is as follows. In section \ref{s2}
we review the transfer matrix method for description of the
multilayer films. In section \ref{s3} we introduce spinor map
between $ SL(2,\mathbb{C}) $ and $ SO(1,3) $, and we draw analogy
between an EM field which passes through a multilayer film and
transformation of coordinates under the Lorentzian group. In
addition, two special cases, a multilayer, that is, 
quarter-wave films and a multilayer with very low thickness films, are
studied in detail. In section \ref{s4}, by using the multilayer
films, we suggest an experimental set-up for investigation of two kinematical
properties, that is, time delay  and length contraction as some of  predictions of
the special relativity. In section \ref{s5} we introduce an analogy
between the Rindler space, as a curved space, and a suitable
multilayer film. Finally, a summary and conclusion  remarks are
given in section \ref{s6}.
\section{Matrix theory for description of the Multilayer films}\label{s2}
\begin{figure}
   \begin{center}
    \includegraphics[angle=0,width=.4\textwidth]{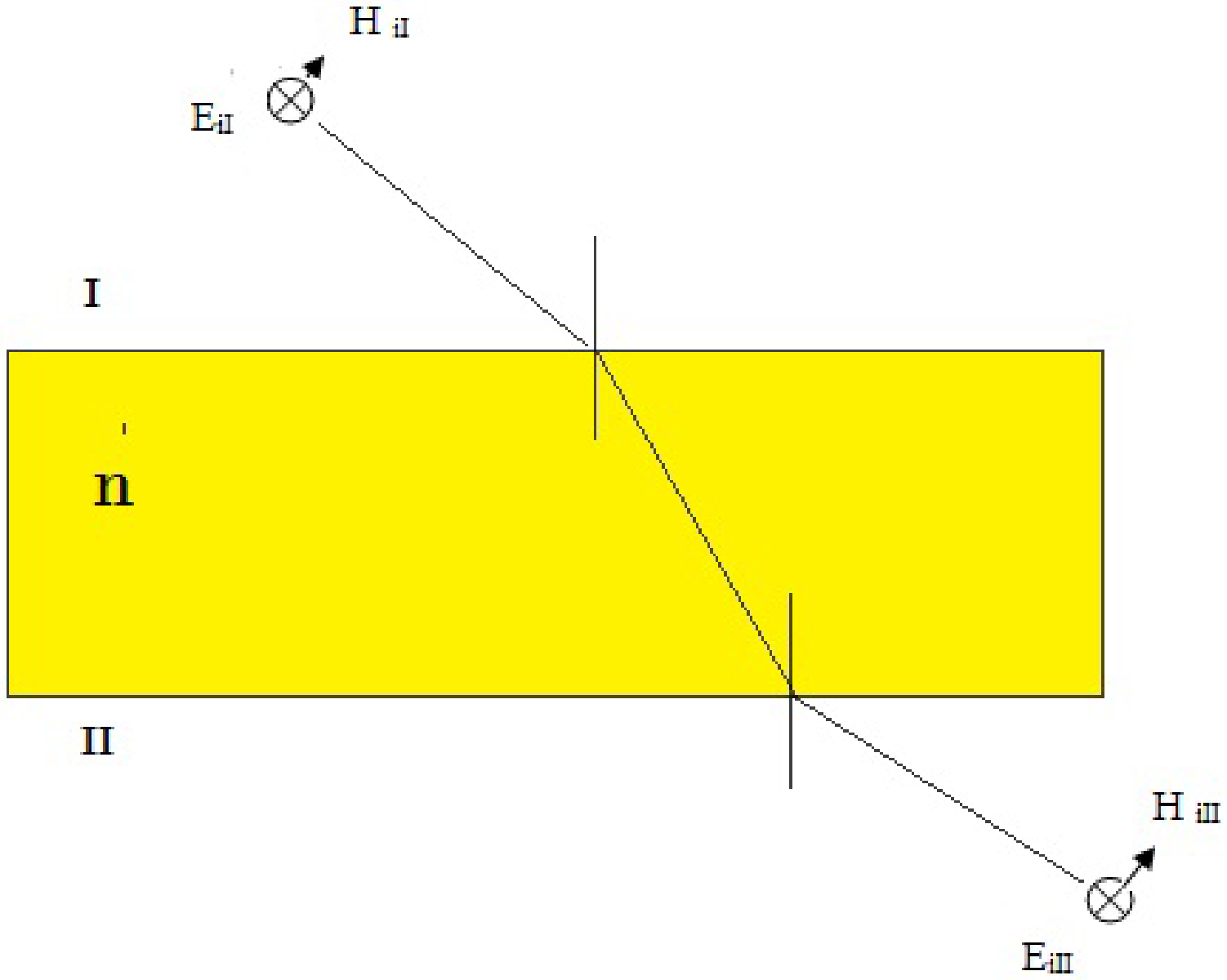}
   \caption{ Scheme of the input and output field in a thin film.}\label{fig1}
\end{center}
 \end{figure}

We consider an electromagnetic radiation falling onto a
structure consisting of a thin film with uniform thickness $ t $
and index of refraction $ n $. This isolated (homogeneous and
isotropic) film, is embedded between two media, 
$I$ and $II$, with indexes of
refraction, $ n_{I} $ and $n_{II} $, respectively. We consider a 
monochromatic, linearly polarized plane falling wave which is 
injected into the thin film from the above
with amplitude $E_{I}$ and $ H_{I} $. The field after passing the end layer
is characterized by $ E_{II} $ and $ H_{II} $, as it is
shown in Figure (\ref{fig1}). We use transfer matrix method which
is a method used in optics  to analyse the
propagation of electromagnetic  through a
stratified (layered) medium \cite{bor,hand}.\\
\indent The transfer-matrix method is based on the fact that, according to
Maxwell's equations, there are simple continuity conditions for
tangential component of  electric field and normal component of 
magnetic field  across boundaries.

 In addition, this method can be developed to the multilayer films.
\cite{hand}. For a stack with $ m $ layers, the transfer-matrix can
be defined as a product of matrices given by
 \begin{equation} \label{1.3}
  \mathcal{M}=M_{m}M_{m-1}\cdots M_{1},
 \end{equation}
where the transfer-matrix of the $j^{th}$ layer, $ M_{j} $, is given by
\begin{equation}\label{12.1}
 M_{j}=\left( \begin{array}{cc}
 \cos \delta_{j} & i \frac{\sin \delta_{j}}{\eta_{j}}\\ i\eta_{j} \sin \delta_{j} &
 \cos \delta_{j}
\end{array} \right),
\end{equation}
where $ \delta_{j}=\frac{2 \pi}{\lambda}\h t_{j} \h n_{j}\cos(\theta_{j}) $
and $ \eta_{j} $ for $s$-polarization and $ p $-polarization is
characterized by $ \eta_{j}^{s}=n_{j} \cos( \theta_{j}) $ and $\eta_{j}^{p}=
n_{j} / \cos( \theta_{j}) $ respectively \cite{hand}. 
It is obvious that the angle $ \theta_{j} $ is related to the
angle of incidence $ \theta_{0} $ by Snell's law, that is,
 \begin{equation}\label{1.4}
  n_{j} \sin \theta_{j}=n_{0} \sin \theta_{0} .
 \end{equation}
\section{Lorentzian interpretation of the transfer matrix}\label{s3}
\begin{figure}
   \begin{center}
    \includegraphics[angle=0,width=.4\textwidth]{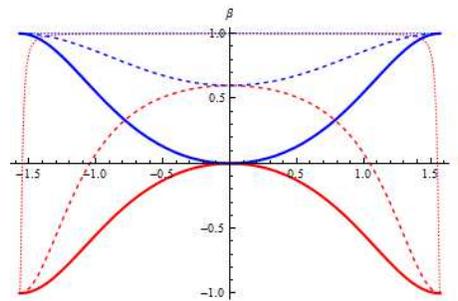}
  \caption{(Colour online) The variation of $ \beta $, for $ s $
  and $p$-polarization,
   with respect to $ \theta $ for different values of the index of refraction, $ n=1 $ are plotted by red and blue line,
   $ n=2 $ by red and blue dashing line and $ n=10 $ by red and blue dot line, for $s$ and $p$-polarization, respectively.}\label{fig3}
   \end{center}
 \end{figure}

\begin{figure}
  \begin{center}
   \includegraphics[angle=0,width=.4\textwidth]{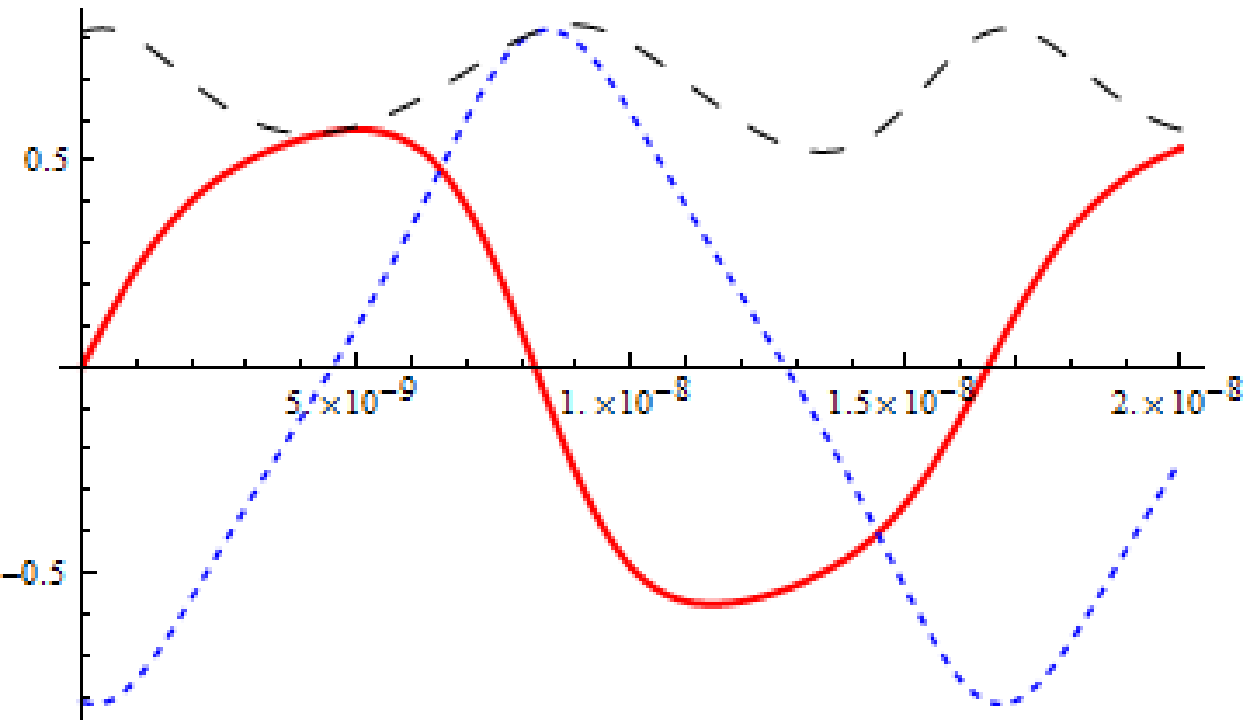}
   \caption{(Colour online)  The variation of $ \beta_{2}^{\prime} $ by red line,
   $ \beta_{3}^{\prime} $ by blue dot line, and $ |\beta^{\prime} | $
  by black dashing line with respect to $ t $ for the fixed index of refraction,
 $ n=2.419 $, for $ s $-polarization.}\label{fig5}
 \end{center}
\end{figure}

\begin{figure}
  \begin{center}
     \includegraphics[angle=0,width=.4\textwidth]{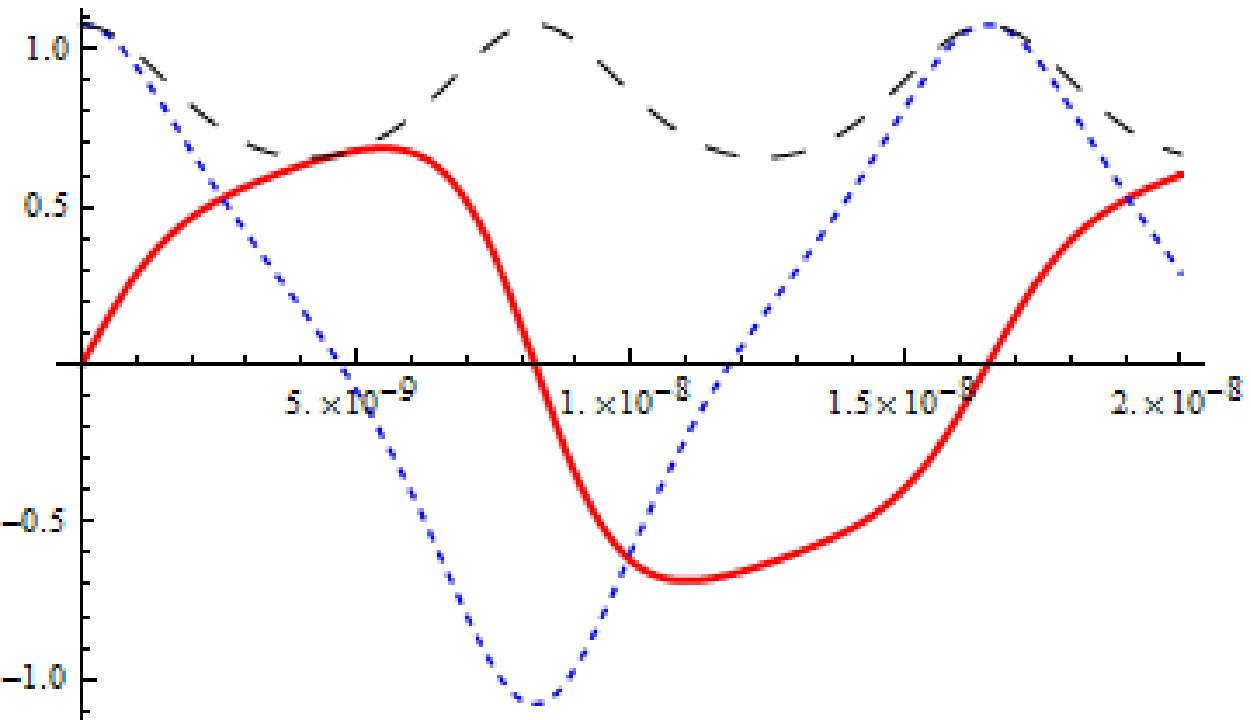}
  \caption{(Colour online) The variation of $ \beta_{2}^{\prime} $ by red line,
   $ \beta_{3}^{\prime} $ by blue dot line, and $ |\beta^{\prime} | $
  by black dashing line with respect to $ t $ for the fixed index of 
  refraction, $ n=2.419 $,
  for $ p $-polarization}\label{fig6}
  \end{center}
 \end{figure}

In this section, we use the spinor map, from $ SL(2,\mathbb{C}) $,
onto the Lorentz group, $ SO(1,3)$. We define a coherence matrix
\cite{bor} as follows:
\begin{equation} \label{2.1}
\mathcal{E}=\left( \begin{array}{cc}
|E|^{2} & E^{*} H\\ E H^{*} & |H|^{2}
\end{array} \right),
\end{equation}
for both $I$ and $II$, so that $\mathcal{E}_I$ and $\mathcal{E}_{II} $ are related acording to the Eq. (\ref{0.4}).
We explicitly define the coordinates $x^{\mu}$, by using equation (\ref{0.5}),
\begin{eqnarray} \label{2.4}
x^{0}&=&\frac{1}{2} (|E|^{2}+|H|^{2}), \hspace{.25cm} x^{1}=\text{Re}
(E^{*}H),\n\\
x^{2}&=& \text{Im}(E^{*}H), \hspace{1cm} x^{3}=\frac{1}{2}
(|E|^{2}-|H|^{2}).
\end{eqnarray}
\indent It is worth to note that these coordinates  behave as a light-like interval: 
\begin{equation} \label{2.11}
(x^{0})^{2}-(x^{1})^{2}-(x^{2})^{2}-(x^{3})^{2}=0,
\end{equation}
\indent Therefore, the conjugation relation (\ref{0.4}) induces a transformation on 
the variables $ x^{\mu}$ in the form of equation (\ref{2.5}).\\
\indent However, we can decompose the transfer matrix of a thin film,
according to equations (\ref{1.3}) and (\ref{12.1}) with $j=1$, as
\begin{eqnarray} \label{2.7}
M=\mathcal{B}^{-1}\ \mathcal{R}\ \mathcal{B}
=\left( \begin{array}{cc}
\frac{1}{\sqrt{\eta}} &0\\ 0 & \sqrt{\eta}
\end{array} \right) \left( \begin{array}{cc}
\cos\delta &i \sin \delta\\ i \sin
\delta & \cos \delta
\end{array} \right) \left( \begin{array}{cc}
\sqrt{\eta} &0\\ 0 & \frac{1}{\sqrt{\eta}}
\end{array} \right).
\end{eqnarray}
Thus, under the homomorphism (\ref{2.6}), $ \mathcal{B}$
generates a boost of velocity $ \beta=v $ (in natural units, where $c=1$), while 
$\mathcal{R} $ induces a pure spatial rotation.\\
\indent To gain physical insights, we write explicitly  the equation
(\ref{2.6}) for
 $ \mathcal{B} $ and $ \mathcal{R} $. Under the homomorphism (\ref{2.6}), $
 \mathcal{R} $ generates  the following matrix
\begin{eqnarray} \label{2.8}
T(\mathcal{R})\equiv R(2\delta)= \left(  \begin{array}{cccc} 1&
0&0&0\\0&1&0&0\\0&0&\cos 2\delta & \sin 2\delta \\ 0&0&
-\sin2\delta & \cos 2\delta\end{array} \right),
\end{eqnarray}
which is a rotation in the plane $x^{2}-x^{3}$ of angle twice the
phase of the transmission coefficient.\\
\indent Furthermore, diagonal matrix $ \mathcal{B}$, under the
homomorphism (\ref{2.6}), generates a boost in the plane $
x^{0}-x^{3} $ \cite{mis1},
\begin{eqnarray} \label{2.9}
T(\mathcal{B})\equiv L(0,0,\beta)
=\left(  \begin{array}{cccc}
\cosh \zeta & 0&0&\sinh \zeta\\0&1&0&0\\0&0&1 & 0 
\\ \sinh \zeta & 0 & 0 & \cosh \zeta
\end{array} \right),
\end{eqnarray}
where rapidity $  \zeta $ is obtained from the relations
\begin{eqnarray} \label{2.10}
\cosh \zeta =\frac{1}{2} (\eta+\frac{1}{\eta}),\\
\sinh \zeta =\frac{1}{2} (\eta-\frac{1}{\eta}),
\end{eqnarray}
and velocity $ \beta $ is defined by
\begin{equation}\label{2.11}
\beta= \tanh \zeta= \frac{(\eta-\frac{1}{\eta})}
{(\eta+\frac{1}{\eta})}.
\end{equation}
\indent In Figure (\ref{fig3}), we have plotted the
variation of $ \beta $ with respect to the $ \theta $ for different
values of the index of refraction, $ n $, when the
electromagnetic field have $s$-, and $ p $-polarization. It is seen that  for
every values of the index of refraction, $\beta$ has maximum or minimum in $\theta=0$.  In
addition, for a fixed
$  \theta$, the $ \beta $ is increased by increasing the refraction index,
for both $s$-, and $p$-polarization. However, for the large values of the
refraction index, $ n\gg 1 $, for both polarization, the result is the same, 
as shown in Figure (\ref{fig3}). Finally, by using equations (\ref{2.7}), (\ref{2.8})
and (\ref{2.9}) we obtain $ T(M) $ as
\begin{eqnarray}\label{2.10.1}
T(M)=L(0,0,-\beta) R(2\omega t) L(0,0,\beta)
\end{eqnarray}
where  $ \omega=\delta/t $. Also, with some calculations,
we can rewrite equation (\ref{2.10.1}), in the following form \cite{sat,corn}. 
\begin{eqnarray}\label{2.100.2}
T(M)=R(2\omega t) L\big(0, \beta \sin(2\omega t),-\beta \cos(2\omega t)\big) 
                 L(0,0,\beta).
\end{eqnarray}
In addition, by using Einstein  rule for the composition of velocities \cite{un},
\begin{equation}\label{3.2}
\vec{\beta}_{1}\oplus \vec{\beta}_{2}= \frac{1}{1+\vec{\beta}_{1}
\vec{\beta}_{2}} \Big\lbrace \vec{\beta}_{1}+\vec{\beta}_{2}+
\frac{\gamma_{\beta_{1}}}{1+\gamma_{\beta_{1}}}\big(\vec{\beta}_{1}\times
(\vec{\beta}_{1}\times\vec{\beta}_{2}) \big)\Big\rbrace,
\end{equation}
we can write $ T(M) $ as a rotation and a boost,
\begin{equation}\label{2.200}
T(M)=R(2\omega t) L(0, \beta_{2}^{\prime},\beta_{3}^{\prime}),
\end{equation}
where
\begin{eqnarray}
\beta_{2}^{\prime}&=&\frac{\beta \sin(2\omega
t)}{1-\beta^{2}\cos(2\omega t)} [1-\frac{\beta^{2}\cos(2\omega
t)}{1+\sqrt{1-\beta^{2}}}],\n\\
\beta_{3}^{\prime}&=&-\frac{\beta }{1-\beta^{2}\cos(2\omega t)}
 \Big( \cos(2\omega t)+\frac{\beta^{2}\sin^{2}(2\omega t)}
{1+\sqrt{1-\beta^{2}}}\Big).
\end{eqnarray}
\indent In Figures (\ref{fig5}) and (\ref{fig6}), we have plotted the
variations of $\beta_{2}^{\prime} $, $ \beta_{3}^{\prime} $ and
$|\beta^{\prime}|=\sqrt{\beta_{2}^{\prime 2}+\beta_{3}^{\prime 2}
} $ with respect to $ t $ for $ n=2.00049 $ and $ \theta =\pi /9
$, for $ s $-polarization and $ p $-polarization, respectively. As
it is seen, $ \beta_{2}^{\prime} $, $\beta_{3}^{\prime} $ and $
|\beta^{\prime}| $ depend periodically on $ t $.\\ 
\indent We now turn to the multi-layer stack with $ m $ films. By employing the
method which is used to obtain equation (\ref{2.100.2}),  and by
using equation (\ref{2.200}), we can write $T(\mathcal{M}) $ as
\begin{eqnarray}\label{3.3}
T(\mathcal{M})=R(2\Omega t) L(0,
\mathcal{B}_{2}^{(m)},\mathcal{B}_{3}^{(m)})\cdots
L(0,\mathcal{B}_{2}^{(1)},\mathcal{B}_{3}^{(1)}),
\end{eqnarray}
where $\Omega=\sum_{i=1}^{m} \omega^{(i)}$ and
\begin{eqnarray}\label{3.5}
\mathcal{B}_{2}^{(i)}&=&\beta_{2}^{(i)}\cos(2\Omega^{(i)}t^{(i)})-
\beta_{3}^{(i)}\sin(2\Omega^{(i)}t^{(i)}),
\\\mathcal{B}_{3}^{(i)}&=&\beta_{2}^{(i)}\sin(2\Omega^{(i)}t^{(i)})+\beta_{3}^
{(i)}\cos(2\Omega^{(i)}t^{(i)}),
\end{eqnarray}
with $ \Omega^{(i)}=\sum_{j=1}^{i-1} \omega^{(j)} $ and
\begin{eqnarray}\label{3.6}
\beta_{2}^{(i)}&=&\frac{\beta^{i} \sin(2\omega^{i} t^{i})}
{1-(\beta^{i})^{2}\cos(2\omega^{i} t^{i})}
[1-\frac{(\beta^{i})^{2}\cos(2\omega^{i} t^{i})}{1+\sqrt{1-(\beta^{i})^{2}}}],\\
\beta_{3}^{(i)}&=&-\frac{\beta^{(i)} }{1-(\beta^{(i)})^{2}\cos(2\omega^{(i)}
t^{(i)})}
\Big( \cos(2\omega^{(i)} t^{(i)})+\frac{(\beta^{(i)})^{2}\sin^{2}(2\omega^{(i)}
t^{(i)})}{1+\sqrt{1-(\beta^{(i)})^{2}}}\Big).
\end{eqnarray}
Also, by using Einstein addition rule (\ref{3.2}), we can rewrite equation (\ref{3.3}) as
\begin{equation}\label{3.7}
T(\mathcal{M})=R(2\Omega t) L(0,
\mathcal{B}_{2}^{(\mathcal{M})},{\mathcal{B}_{3}^{(\mathcal{M})}}),
\end{equation}
where $ \mathcal{B}_{2}^{(\mathcal{M})} $ and $
\mathcal{B}_{3}^{(\mathcal{M})} $ are given by applying the rule (\ref{3.2}) consecutively.\\
\subsection{A multilayer with quarter-wave films}
In this section, we consider a multilayer with quarter-wave
films \cite{pedr}, that is $ t_{j} n_{j}=\lambda/4 $ and assume
that $\theta_{0}=0 $, so that $\cos \delta_{j} = 0$ and $\sin
\delta_{j} = 1$. As an example, for a stack consists of two
layers, matrix $ M$, is given by,
\begin{eqnarray} \label{201.1}
M=\left(  \begin{array}{cc}
-\frac{n_{1}}{n_{2}} & 0\\0&-\frac{n_{2}}{n_{1}}
\end{array} \right),
\end{eqnarray}
which under the homomorphism (\ref{2.6}), generates a boost as \cite{mis1},
\begin{eqnarray} \label{20.9}
T(M)\equiv L(0,0,\beta_{qw}^{(2)})
=\left(  \begin{array}{cccc}
\cosh \zeta_{qw}^{(2)} & 0&0&\sinh \zeta_{qw}^{(2)}
\\0&1&0&0\\0&0&1 & 0 \\ \sinh
\zeta_{qw}^{(2)} & 0 & 0 & \cosh \zeta_{qw}^{(2)}
\end{array} \right),
\end{eqnarray}
where velocity $ \beta_{qw}^{(2)} $ is given by
\begin{equation}
\beta_{qw}^{(2)}= \tanh \zeta_{qw}^{(2)}= \frac{1-(\frac{n_{2}}{n_{1}})^{4}}
{1+(\frac{n_{2}}{n_{1}})^{4}},
\end{equation}
and rapidity $  \zeta_{qw}^{(2)} $ is obtained from the relations
\begin{eqnarray} \label{20.10}
\cosh \zeta_{qw}^{(2)} =\frac{1}{2} \Big((\frac{n_{1}}{n_{2}})^{2}+(\frac{n_{2}}
{n_{1}})^{2}\Big),\\
\sinh \zeta_{qw}^{(2)} =\frac{1}{2} \Big((\frac{n_{1}}{n_{2}})^{2}-(\frac{n_{2}}
{n_{1}})^{2}\Big).
\end{eqnarray}
\indent In general, for a stack consists of $ 2m $ quarter-wave layers,
in which the thickness of $ j^{th} $ layer is $ t_{j}=\frac{\lambda}{4
n_{j}},\ j=1,2,\cdots,2m $,
after some simple calculation, we can achieve $ T(M_{2m}) $ as,
\begin{eqnarray} \label{21.9}
T(M_{2m})\equiv L(0,0,\beta_{qw}^{(2m)})
=\left(  \begin{array}{cccc}
\cosh \zeta_{qw}^{(2m)} & 0&0&\sinh \zeta_{qw}^{(2m)}\\0&1&0&0\\0&0&1 & 0 \\ \sinh \zeta_{qw}^{(2m)} & 0 & 0 & \cosh \zeta_{qw}^{(2m)}
\end{array} \right),
\end{eqnarray}
where velocity $ \beta_{qw}^{(2m)} $ is calculated by the rule (\ref{3.2}).  As an example for $2m=4$, we have
\begin{equation}\label{4.100}
\beta_{qw}^{(4)}=\tanh \zeta_{qw}^{(4)}=\frac{\beta_{1}+\beta_{2}}{1+\beta_{1}\beta_{2}} .
\end{equation}
\subsection{A multilayer with very low thickness} \label{thin}
In this section we study a multilayer, with $ m $ layers,  in such a way that $ \eta^{(i)}$'s are near to the unit, $ \eta^{(i)}  \simeq 1$. Thus, by using equation (\ref{2.11}) and keeping only first-order term of $ \eta^{(i)}$, we obtain  $\beta^{(i)}=-1+\eta^{(i)} $ and therefore
 the Lorentzian  transformations are
replaced by the Galilean transformations.  Thus, by these assumption, we can
rewrite the equation (\ref{3.7}) as the following relation: 
\begin{equation}\label{4.1}
T(\mathcal{M})=R(2\Omega t) L(0,
\mathcal{B}_{2}^{(\mathcal{M})},{\mathcal{B}_{3}^{(\mathcal{M})}}),
\end{equation}
where 
\begin{eqnarray}\label{3.10}
\mathcal{B}_{2}^{(\mathcal{M})}&=&\sum_{i=1}^{m}\beta^{(i)}\sin[2(\omega^
{(i)}+\Omega^{(i)}) t^{(i)}],\n\\
 \mathcal{B}_{3}^{(\mathcal{M})}&=&\sum_{i=1}^{m}\beta^{(i)}\cos[2(\omega^
{(i)}+\Omega^{(i)}) t^{(i)}].
\end{eqnarray}
\indent In addition, when we assume $ t $ is very thin, from equation
(\ref{2.10.1}), by keeping only first-order terms in $ t $, the rotation
$ R(2 \omega\h t) $ appears as a pure boost, $ L(0,0,\beta) $,
which is followed by another pure boost in another direction, $ L(0,2a\h
dt,0) $,  so that the result is like a Thomas precession.
That is, we can write \cite{mis1,visser2},
\begin{equation}\label{3.8}
R(2\omega\h dt)= L(0, -2a\h dt, -\beta)  L(0,2a\h dt,0)
L(0,0,\beta),
\end{equation}
where $ \omega=\frac{2 \pi}{\lambda}\h n \cos \theta_{II} $ and $
a=-\frac{\omega}{\gamma}\h $ is the acceleration along the second
axis, $ x^{2} $.
\section{Experimental Set-up of Time Delay and Length Contraction by Thin 
Films}\label{s4}
We can now propose some specific experiments bases on
the properties discussed in the preceding sections. The basic
experimental set-up consists essentially of a monochromatic plane
wave, linear polarized ($ s $ or $ p $) with the specific values
of $|E|$ and $|H|$.\\ 
\indent We use a stack consists of $ 2 $
quarter-wave layers. According to the equation (\ref{20.9}), the coordinates $x^{1}$
and $x^{2}$ under the action of transformation matrix remain invariant. 
Therefore, it is possible to consider $T(\mathcal{M)}$ as a boost which belongs to $SO(1,1)$:
\begin{eqnarray} \label{5.1}
\left( \begin{array}{c}x_{out}^{0}\\ x_{out}^{3}\end{array}
\right)=\frac{1}{2} (\frac{n_{1}}{n_{2}})^{2}\left(
\begin{array}{cc} 1+(\frac{n_{2}}{n_{1}})^{4}&
1-(\frac{n_{2}}{n_{1}})^{4}\\1-(\frac{n_{2}}{n_{1}})^{4}&1
+(\frac{n_{2}}{n_{1}})^{4}\end{array} \right)\left(
\begin{array}{c}x_{in}^{0}\\ x_{in}^{3}\end{array} \right).
\end{eqnarray}
\indent To show how the experimental set-up works, let us suppose that
the electromagnetic fields, $E_{in}$ and $H_{in}$, fall onto the 
multi-layer. From equation (\ref{2.4}) we have
\begin{equation}\label{5.2}
x_{in}^{0}=\frac{1}{2}(|E_{in}|^{2}+|H_{in}|^{2}), \hspace{.25cm}
x_{in}^{3}=\frac{1}{2}(|E_{in}|^{2}-|H_{in}|^{2}),
\end{equation}
and
\begin{eqnarray}\label{5.3}
x_{out}^{0}&=&\frac{1}{2}(|E_{out}|^{2}+|H_{out}|^{2})
=\frac{1}{2}(\frac{n_{1}}{n_{2}})^{2}\h \{|E_{in}|^{2}+(\frac{n_{2}}
{n_{1}})^{4}|H_{in}|^{2}\}, \n\\
 x_{out}^{3}&=&\frac{1}{2}(|E_{out}|^{2}-|H_{out}|^{2})
=\frac{1}{2}(\frac{n_{1}}{n_{2}})^{2}\h\{|E_{in}|^{2}-(\frac{n_{2}}
{n_{1}})^{4}|H_{in}|^{2}\}.
\end{eqnarray}
These equations clearly show that there are many possible
experimental methods to choose the input fields in such a way that
two different points in a reference frame may be appear in the
other frame as either with the same temporal or same spatial
coordinate.\\
\indent We consider two experimental set-ups with different inputs 
$E_{in(j)}$ and $H_{in(j)}$ and corresponding to outputs $E_{out(j)}$
 and $H_{out(j)}$ ,where $j=1,2$. 
Thus, it is easy to show that
$x_{out(1)}^{0}=x_{out(2)}^{0}$ when
\begin{equation}\label{5.4}
|E_{in(2)}|^{2}-|E_{in(1)}|^{2}=(\frac{n_{2}}{n_{1}})^{4}\h\{|H_{in(1)}|^{2}-|
H_{in(2)}|^{2}\},
\end{equation}
and $x_{out(1)}^{3}=x_{out(2)}^{3}$ when
\begin{equation}\label{5.5}
|E_{in(2)}|^{2}-|E_{in(1)}|^{2}=(\frac{n_{2}}{n_{1}})^{4}\h\{|H_{in(2)}|^{2}-|
H_{in(1)}|^{2}\}.
\end{equation}
\indent By using equation (\ref{5.4}), the first set-up can be used to 
demonstrate the length contraction, 
\begin{equation}
l_{out}\equiv
x_{out(2)}^{3}-x_{out(1)}^{3}=\frac{1}{\gamma}(x_{in(2)}^{3}-x_{in(1)}^{3}).
\end{equation}
In order to put forward the phenomenon of the time dilation, let
us consider another experimental set-up in which  $E_{in(1)}$, $H_{in(1)}$,
$E_{in(2)}$ and $H_{in(2)}$ satisfy the equation
(\ref{5.5}). Therefore, measurement of $\Delta
t_{out}\equiv x_{out(2)}^{0}-x_{out(1)}^{0}$ may be regarded as a
time interval measurement. Therefore, we obtain
\begin{equation}
\Delta t_{out}\equiv
x_{out(2)}^{0}-x_{out(1)}^{0}=\frac{1}{\gamma}(x_{in(2)}^{0}-x_{in(1)}^{0}).
\end{equation}
\indent In summary, we have proposed a series of specific experiments
by using a multilayer, which contains two films and the thickness of
their films are given by $t_{j} n_{j}=\lambda/4$, that clearly
demonstrate the most important predictions of the Lorentz
kinematics.
\section{A Multi-layer as a Rindler Space}\label{s5}
\begin{figure}
  \begin{center}
  \includegraphics[angle=0,width=.4\textwidth]{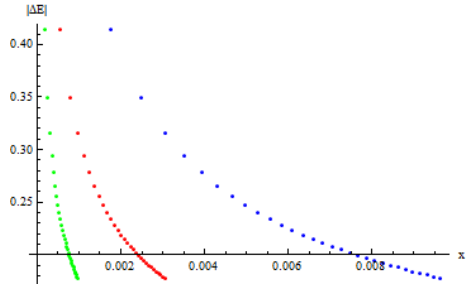}
  \caption{(Colour online) The variation of $ \Delta E $ with respect to $ x=\sqrt{\frac{\lambda}{2g}j} $
  with $g=1$ and $2m=60$ for $ \lambda=0.06199\times 10^{-6} $, $ \lambda=0.06199\times 10^{-5} $
  and $ \lambda=0.06199\times 10^{-4} $ by 
  green ( left), red ( middle) and blue (right)  points, 
  respectively.
  }\label{fig18}
  \end{center}
 \end{figure}

Maxwell's equations in general coordinates are written as
\cite{leo}
 \begin{eqnarray}\label{2.21}
(\sqrt{g}\h g^{ij}
E_{j})_{,i}=\frac{\sqrt{g}\h\rho}{\varepsilon_{0}},\hspace{.5cm}
(\sqrt{g}\h g^{ij}B_{j})_{,i}=0,
\end{eqnarray}
and
\begin{eqnarray}\label{2.22}
\left[ ijk \right] E_{k,j}=\dfrac{\partial (\sqrt{g}\h g^{ij}B_{j})}{\partial t},\hspace{.5cm}
 \left[ ijk \right] H_{k,j}=\frac{1}{c^{2}}\dfrac{\partial (\sqrt{g}\h g^{ij}E_{j})}{\partial t}+\mu_{0}\sqrt{g}\h j^{i},
\end{eqnarray}
where permutation symbol, $ \left[ ijk \right] $, is defined by
\begin{eqnarray}\label{2.23}
\left[ ijk \right]=\Bigg\lbrace \begin{array}{ccc}
+1 ,& \text{if}\ ijk \hspace{.1cm} $ is an even permutation of $ 123,\\
-1 , & \text{if}\ ijk \hspace{.1cm}$ is an odd permutation of $ 123,\\
 0 , & $otherwise$.
 \end{array}
\end{eqnarray}
and the Einstein summation convention is used. 
As it is well known in Optics, the Maxwell's equations in
empty-curved space can be replaced by Maxwell's equations in the flat space which
is filled by dielectric media with the following parameters \cite{leo},
 \begin{equation} \label{2.24}
\varepsilon^{ij}=\mu^{ij}=\pm \sqrt{g}\h g^{ij}.
\end{equation}
Therefore, we can see a correspondence between geometry 
and matter, as in the general relativity. Thus, in
this case, the index of refraction is obtained by \cite{leo}
\begin{equation}\label{2.267}
n^{2}=\left( \begin{array}{ccc}
\varepsilon_{11}\varepsilon_{22}
-\varepsilon_{12}\varepsilon_{21}\hspace{.2cm} &\varepsilon_{13}\varepsilon_{32}
-\varepsilon_{12}\varepsilon_{33}\hspace{.2cm} &\varepsilon_{12}\varepsilon_{23}
-\varepsilon_{13}\varepsilon_{22}\\
\varepsilon_{23}\varepsilon_{31}-\varepsilon_{21}\varepsilon_{33}\hspace{.2cm} &
\varepsilon_{11}\varepsilon_{33}-\varepsilon_{13}\varepsilon_{31}\hspace{.2cm} &
\varepsilon_{13}\varepsilon_{21}-\varepsilon_{13}\varepsilon_{23}\\
\varepsilon_{21}\varepsilon_{12}-\varepsilon_{22}\varepsilon_{31}\hspace{.2cm} &
\varepsilon_{12}\varepsilon_{31}-\varepsilon_{11}\varepsilon_{32}\hspace{.2cm} &
\varepsilon_{11}\varepsilon_{22}-\varepsilon_{12}\varepsilon_{21}
\end{array}\right)
\end{equation}
\indent Now, if we consider the metric,
 \begin{equation}\label{2.22}
ds^{2}=-dt^{2}+N^{2}(x)(dx^{2}+dy^{2}+dz^{2}),
\end{equation}
by using equations (\ref{2.24}) and (\ref{2.267}), the index of
refraction is given by $ n(x)=[N(x)]^{-1} $.\\
\indent On the other hand, the principle of equivalence make the
gravitational field connect to the acceleration. 
Therefore, to study gravitational effects, we can
generalize concept of Lorentz transformations so as to include
observers who are moving with a constant acceleration with
respect to an inertial frame. This metric is called  Rindler
metric and described as \cite{pad},
 \begin{equation}\label{2.25}
ds^{2}=-(g\ x)^{2}dt^{2}+dx^{2}+dy^{2}+dz^{2}.
\end{equation}
It is easy to see that the metrics (\ref{2.22}) and (\ref{2.25}) are conformal. 
Accordingly, by using the equation (\ref{2.267}), we obtain  the refractive index 
of the analogue media with the Rindler space as the following relation:
\begin{equation}\label{2.26}
n= g \ x.
\end{equation}
Thus, for analogue Rindler space we assume a multilayer, for  which
the refractive index of the $ j^{th} $ layer is characterized by,
\begin{equation}\label{4.5}
n_{j}= g(j-\frac{1}{2})t.
\end{equation}
However, up to the first order approximation in $\beta^{(i)}$
and by using equation (\ref{4.5}), for $ s $- and
$p$-polarization, when $ \theta_{0}=0 $, we can rewrite equation (\ref{4.1}) 
as,
\begin{equation}
T(\mathcal{M})=L(0,\mathbb{B}_{2}, \mathbb{B}_{3})
\end{equation}
where
\begin{eqnarray}
\mathbb{B}_{2}&=& \sum_{i=0}^{m}\frac{(g(j-\frac{1}{2})t)^{2}-1}{(g(j-\frac{1}
{2})t)^{2}+1}
\sin(\frac{2\pi}{\lambda} t^{2}i^{2}),
\end{eqnarray}
and
\begin{eqnarray}
\mathbb{B}_{3}&=&
\sum_{i=0}^{m}\frac{(g(j-\frac{1}{2})t)^{2}-1}{(g(j-\frac{1}{2})t)^{2}+1}
\cos(\frac{2\pi}{\lambda} t^{2}i^{2}).
\end{eqnarray}
\indent Also, we can consider a stack consists of the quarter-wavelength
layers, i.e., $ t_{j}n_{j}=\lambda/4 $. Therefore, by using the 
equation (\ref{2.26}), the thickness of $ j^{th} $ layer is
\begin{equation}
t_{j}=\sqrt{\frac{\lambda}{2g}+x_{j}^{2}}-x_{j},
\hspace{.5cm}x_{j}=\sum_{i=1}^{j-1}t_{i}.
\end{equation}
After some calculation, we can obtain thickness of the $
j^{th} $ layer as,
\begin{equation}\label{4.101}
t_{j}=\sqrt{\frac{\lambda}{2g}}(\sqrt{j}-\sqrt{j-1}),
\end{equation}
and the index of refraction of this layer is obtained by
\begin{equation}\label{4.102}
n_{j}=\sqrt{\frac{\lambda g}{8}} (\sqrt{j}+\sqrt{j-1}).
\end{equation}
We can now construct an analogy between the Rindler space and a multilayer with
$ 2m $ quarter-wavelength layers, in which the thickness of $j^{th}$ layer
is given by equation (\ref{4.101}) and its refraction index 
is given by equation (\ref{4.102}). Accordingly we obtain
,
\begin{equation}\label{Multilyer}
E_{Multilyer}=\frac{\sqrt{j+1}+\sqrt{j}}{\sqrt{j}+\sqrt{j-1}} E_{0}
\end{equation}
\indent In Refs. \cite{30,31}, electrodynamics in the Rindler space had
been studied by using a phenomenological approach. Accordingly,
the component of the electric field, near the horizon, $x=0$, reads as follow:
\begin{equation}\label{rin}
E_{\text{Rindler}}(x)=E_{0}\ K_{i 2\pi/\lambda}(\frac{2\pi}{\lambda}x),
\end{equation}
where $K_{i 2\pi/\lambda}(2\pi x/\lambda)$ is the modified Bessel
function of order $i2\pi/\lambda$.\\
\indent   Figure (\ref{fig18}) displays a comparison between the output values
of the electric fields are given by equation  (\ref{Multilyer})  and equation (\ref{rin}), i.e.,  
\[\Delta E=\text{Abs}(|E|_{\text{Multilayer}}-|E|_{\text{Rindler}}),\]
 in terms of  $x=\sqrt{\frac{\lambda}{2g}j}$ and for different
values of $\lambda$. As it is seen, by increasing  $x$
(increasing the number of layers), $\Delta E$ approaches to zero.
In addition, $\Delta E$ is decreased more rapidly by decreasing
the $\lambda$.
\section{conclusion} \label{s6}
This paper is an example of analogue  models of Einstein Relativity. 
Firstly, we have used a geometrical tool to analyze a multilayer film
in a concise way that is closely related to the special
relativity. Then, this method was made an analogy between
multilayer stack and curved space in the general relativity. In
fact, in the first step, we have seen how the transfer matrix 
can be interpreted as multiplication of a boost, a
rotation and another boost.\\
\indent In addition, in this case, we have suggested an experimental
set-up which tests the most important predictions of the Lorentz
kinematics, i.e., time delay and length contraction. However,
by using this isomorphism, apart from a relativistic presentation
of the topic, which has interest in its own, we suggested an
experimental set-up which is, in theoretical view,  so simple and
understandable, and, in practical level, was so easy to
accomplish, in contrast to the similar experimental set-up
\cite{mon12}.\\
\indent Finally, we have employed analogue method in replacing Maxwell's
equations in empty-curved space to Maxwell's equations in the
flat space where the later  is filled by a media. Then, we have applied this
method to the Rindler frame, which by using principle of
equivalence, is locally equivalent to gravitational fields. This
means that this method makes an analogy between accelerated frame
in general relativity and a suitable multilayer films, with
specific values of the refractive indexes and thickness of the
films.
\section{Acknowledgement}
The authors wish to thank The Office of Graduate Studies of The
University of Isfahan and Shahrekord University for their support.

 \vspace{1cm}



\end{document}